\documentclass[aps, prl, reprint, superscriptaddress]{revtex4-2} 

\usepackage{amsmath, amssymb, 
bm, mathtools, mathrsfs, 
ulem}
\usepackage{gensymb,nccmath} 
\usepackage{hyperref, cleveref, natbib} 
    \hypersetup{
    	colorlinks=true,
    	linkcolor=blue,
    	breaklinks=true,
    	filecolor=green,      
    	urlcolor=cyan,
    	citecolor=magenta
    }
\usepackage{graphicx, xcolor}

\newcommand{\BW}{\mathcal{B}} 
\newcommand{\DeltaComplex}{\mathbf{\Delta}} 
\newcommand{\ket}[1]{|#1\rangle} 
\newcommand{\bra}[1]{\langle#1|} 
\newcommand{\abs}[1]{|#1|}

\newcommand{\overlapW}{\mathsf{W}} 
\newcommand{\hamh}{\mathsf{h}} 
\newcommand{\overlapw}{\mathsf{w}} 
\newcommand{\hamH}{\mathsf{H}} 
\renewcommand{\emph}[1]{\textit{#1}}
\newcommand{\DeltaBCS}{\Delta_{\rm BCS} }
\newcommand{\higgs}{{\rm H} }
\newcommand{\matH}{\textsf{\textbf{H}}}
\newcommand{\matW}{\textsf{\textbf{W}}}

\begin{document}
\title{ Higgs gap modes in  superconducting circuit quantisation}

\author{Yun-Chih Liao}
\email{yunchih.liao@uq.edu.au}
\affiliation{ARC Centre of Excellence for Engineered Quantum Systems}
\affiliation{School of Mathematics and Physics, The University of Queensland,  Brisbane, Queensland 4072, Australia}

\author{Benjamin J. Powell}
\affiliation{School of Mathematics and Physics, The University of Queensland,  Brisbane, Queensland 4072, Australia}

\author{Thomas M. Stace}
\email{stace@physics.uq.edu.au}
\affiliation{ARC Centre of Excellence for Engineered Quantum Systems}
\affiliation{School of Mathematics and Physics, The University of Queensland,  Brisbane, Queensland 4072, Australia}

\date{\today}

\begin{abstract}
We extend a recently developed projective circuit quantisation approach to incorporate  superconducting Higgs modes associated to gap dynamics.  This approach starts from a microscopic fermionic Hamiltonian for mesoscopic superconductors, and projects the system onto a low-energy, multi-determinant `BCS' Hilbert space. 
 We derive  analytical results for the superconducting Higgs mass, `spring' constant, and oscillation frequency of the gap dynamics, which we validate  numerically. %
 We compute  anharmonic corrections to the Higgs frequency for higher excitations of small superconducting islands, and compare our results to previous long-wavelength calculations.
\end{abstract}

\maketitle

The key to understanding superconductivity was the discovery of the complex order-parameter, $\DeltaComplex\equiv\Delta e^{i \phi}$.  Two types of collective modes emerge:  
  Nambu-Goldstone phase modes associated to dynamics of $\phi$, which is the foundational degree of freedom underpinning the theory of superconducting quantum circuits,
and  massive Higgs `gap' modes associated to  dynamics of $\Delta$  \cite{ref:GaugeBosons_mass, ref:superconductivity_quasiprticles&gaudeInvariant, ref:plasmon}, which is comparatively underexplored. 
The microscopic theory of superconductivity was established by Bardeen, Cooper and Schrieffer (BCS)  to compute the equilibrium gap amplitude $\DeltaBCS$ \cite{ref:BCS_superconductivity_theory, ref:RPAinSC, ref:CDW&SC, ref:superconductivity_newMethod-Bogoljubov}, as a central physical quantity.

The frequency of  Higgs modes is predicted in Ginzburg-Landau (GL) mean-field theory to soften as the temperature $T$ approaches the critical temperature,  \mbox{$\omega_\higgs\propto\sqrt{T_c - T}$} \cite{ref:GL}. 
At long wavelength and $T=0$, microscopic  models indicate that \mbox{$\omega_\higgs\approx2\DeltaBCS$} 
 \cite{ref:RabiOsc_timeDepBCSpairing, ref:relaxation&oscillation_OrderParam-FermionicCond, ref:collisionlessRelaxation_SCgap, ref:multifreqOsc, ref:pump-probeResponse_BCSstateDynamics, ref:WignerDistFun_orderParamDynamics,ref:Higgs_dynamics_Floquet}.  
Experimental observations of  Higgs modes in superconductors are notably sparse due to  weak electromagnetic coupling  \cite{ref:review-SCHiggsMode, ref:collectiveHiggsMode&couplingCDW}, with some results for charge-density-waves \cite{ref:SCgapExcitation_RomanScattering, ref:SCgapExcitation_RomanScattering-Bfield}, and terahertz spectroscopy \cite{ref:exp_THz-SCHiggsMode, ref:exp_lightInd-SCHiggsMode, ref:Higgs_disorderedSC}. 
 Bulk superconducting experiments, using non-equilibrium anti-Stokes Raman scattering \cite{ref:antiStokeRamanSpectroscopy}, or quantum pathway interference  \cite{ref:AxialHiggsMode},  have reported \mbox{$\omega_\higgs<2\DeltaBCS$}. 

In parallel, superconducting circuits are now important platforms for quantum  processors. The underlying circuit theory quantises the   GL phase angle $\phi\in[-\pi,\pi)$ \cite{https://doi.org/10.1002/cta.2359,ref:2nlinearSCqubit,ref:compact-noncompact},  implicitly assuming a fixed, mean-field   amplitude, $\Delta\!=\!\DeltaBCS$.  Practically, this  has been very successful, but has some conceptual weaknesses \cite{ref:Mizel2024,ref:mypaper_JJquantisation}.  

\begin{figure}[t]
\includegraphics{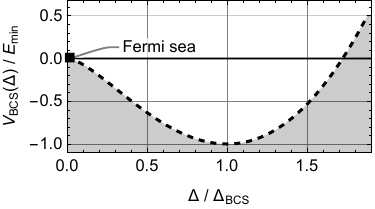}
\caption{
The BCS potential (dashed), given by the diagonal part of the Hamiltonian function \mbox{$\mathsf{V}_{\rm BCS}(\Delta)\equiv\hamH(\varphi=0;\Delta,\Delta)$}.  The energy is minimised at the BCS mean-field  value \mbox{$\Delta=\DeltaBCS$}, where $ \mathsf{V}_{\rm BCS}=-E_{\rm min}=-n\BW e^{-2/\lambda}$. The state at $\Delta=0$ is the zero-temperature Fermi sea.
}
\label{fig:Hdiag}
\end{figure}

In this letter, we extend recent  `first principles' projective approaches that derive quantum  circuit theory directly from a microscopic  Hamiltonian \cite{ref:Mizel2024,ref:mypaper_JJquantisation}, to include  gap dynamics in a unified quantum treatment of both $\phi$ and $\Delta$ in superconducting circuits.  
 Our results yield estimates for the frequency and anharmonicity of gap-mode transitions for small superconducting islands. 
 
We start from a microscopic electronic Hamiltonian for a small, isolated island of superconducting metal consisting of $n$ electronic modes near half-filling,  
\begin{equation}    \label{eqn:Hamiltonian_SC}
    \hat H_\text{el}
=\!
    \sum_{\bm{k}s}\!
    \epsilon_{\bm{k}}   c_{\bm{k}s}^\dagger c_{\bm{k}s}
    -
    |g|^2
    \sum_{\bm{k}\bm{k}'}\!
    c_{\bm{k}\uparrow}^\dagger  c_{-\bm{k}\downarrow}^\dagger
    c_{-\bm{k}'\downarrow}   c_{\bm{k}'\uparrow}+E_F,
\end{equation}
where  $-\BW\!<\!\epsilon_{\bm{k}}\!<\!\BW$ is the single-electron kinetic energy for mode $\bm{k}$ in the  bandwidth $\pm\BW$,   $s\in\{\uparrow,\downarrow\}$, $g$ is the effective electron-electron binding energy,  $c_\bullet^{(\dagger)}$  are  electronic mode creation and annihilation operators, and $E_F=n\BW/2$. 
The low-energy physics is controlled by the dimensionless material parameter \mbox{$\lambda={\abs{g}^2 n}/({2\mathcal{B}})\ll1$}, which is independent of the system size $n$  (since $\abs{g}^2\propto1/n$) \cite{ref:annett}.

We assume that the island is  smaller than the coherence length, so that the low-energy  subspace is spanned by a basis of BCS  states parameterised by a  complex parameter $\DeltaComplex$, \mbox{$\mathcal{H}_{\rm BCS}=\rm{span}\{\ket{\Psi (\DeltaComplex)}\}_{\DeltaComplex \in \mathcal{S}\subset \mathbb{C}}$}, where 
\begin{equation}    \label{eqn:BCSstate_phase-gap}
    \ket{\Psi (\DeltaComplex)}
=
    {\prod}_{\bm{k}}
    (
        u_{\bm{k}}(\Delta)
        +
        v_{\bm{k}}(\Delta)e^{i\phi}
        c_{\bm{k}\uparrow}^\dagger  c_{-\bm{k}\downarrow}^\dagger
    )
    \ket{0},
\end{equation}
 with 
 $u_{\bm{k}}^2=\tfrac12(1 + \epsilon_{\bm{k}} / \sqrt{\epsilon_{\bm{k}} ^2 + \Delta^2})=1-v_{\bm{k}}^2$, $\Delta\geq0$, and $\ket{0}$ is the electronic vacuum   \cite{ref:annett,ref:mypaper_JJquantisation}.   In the projective approach, we define a generalised projector  \mbox{$\hat\Pi=\int_\mathcal{S} d\DeltaComplex  \ket{\Psi (\DeltaComplex)}\bra{\Psi (\DeltaComplex)}$} with which to  project out a low-energy circuit Hamiltonian  
\begin{align}
\hat{\hamH}\equiv\hat\Pi\hat H_\text{el}\hat\Pi 
={\medint{\int_{\mathcal{S}^2}}}  d\DeltaComplex d\DeltaComplex'  \,\hamH(\DeltaComplex,\DeltaComplex') \ket{\Psi(\DeltaComplex)}\bra{\Psi(\DeltaComplex')},
\end{align}  
where \mbox{$\hamH(\DeltaComplex,\DeltaComplex')=\bra{\Psi (\DeltaComplex)}\hat H_\text{el} \ket{\Psi (\DeltaComplex')}$} is  the  Hamiltonian function for the superconducting system.

The BCS basis is normalised,  \mbox{$\langle{\Psi (\DeltaComplex)}|{\Psi (\DeltaComplex)}\rangle=1$} (since $u_{\bm{k}}^2+v_{\bm{k}}^2=1$), but  not  orthogonal, so  $\hat\Pi$ is not idempotent,
\begin{equation}
\hat\Pi^2=\medint{\int_{\mathcal{S}^2}}  d\DeltaComplex d\DeltaComplex'  \,\overlapW(\DeltaComplex,\DeltaComplex') \ket{\Psi(\DeltaComplex)}\bra{\Psi(\DeltaComplex')}\neq\hat\Pi,
\end{equation}
where  \mbox{$\overlapW(\DeltaComplex,\!\DeltaComplex')\!=\!\langle{\Psi (\DeltaComplex)}|{\Psi (\DeltaComplex')}\rangle$} is the  overlap function. 
 
 Electronic eigenstates satisfy \mbox{$\hat H_\text{el}\ket{y}=E\ket{y}$, and} assuming that  low-energy eigenstates  reside within $\mathcal{H}_{\rm BCS}$, i.e.\ $\ket{y}=\int_\mathcal{S} d\DeltaComplex'  \,y(\DeltaComplex') \ket{\Psi (\DeltaComplex')}$, it  follows that   \mbox{$
\int_\mathcal{S} d\DeltaComplex' \,y(\DeltaComplex')  \hat H_\text{el} \ket{\Psi (\DeltaComplex')} =E  \int_\mathcal{S} d\DeltaComplex'  \,y(\DeltaComplex')  |{\Psi (\DeltaComplex')}\rangle \nonumber
$}. This forms a natural superconducting generalisation of the multi-determinant wavefunctions familiar from post-Hartree-Fock methods \cite{fulde}. Projecting over $\bra{\Psi (\DeltaComplex)}$ yields the generalised integral eigenvalue equation
\begin{equation}
{\medint{\int_{\mathcal{S}}}} d\DeltaComplex'\, \hamH(\DeltaComplex,\DeltaComplex')y(\DeltaComplex')=E  {\medint{\int_{\mathcal{S}}}} d\DeltaComplex' \,\overlapW(\DeltaComplex,\DeltaComplex')y(\DeltaComplex').
\label{eqn:geneval}
\end{equation}

The overlap  $\overlapW$ was discussed  in \cite{ref:mypaper_JJquantisation} as a function of the phase coordinate, $\phi$, under the BCS mean-field approximation   \mbox{$|\DeltaComplex|=\DeltaBCS =2\BW e^{-1/\lambda}$} \cite{ref:annett}, i.e.\ assuming \mbox{$\DeltaComplex\in\mathcal{S}=\{\DeltaBCS e^{i\phi}\}_{\phi\in[-\pi,\pi)}$}.  That  earlier analysis  yielded low-energy quantised phase and charge operators $\hat\phi$ and $\hat n$, as well as  standard expressions for quantised superconducting circuit elements which depend on them.  

Here, we extend the projective approach to expand  the support of the gap parameter from \mbox{$\Delta=\DeltaBCS $} to  \mbox{$\Delta\in \mathbb{R}^+$}. 
In the expanded `gap' space, we follow a similar  analysis to \cite{ref:mypaper_JJquantisation},  and compute 
\begin{align} \label{eqn:overlap_phase-gap}
\overlapW (\DeltaComplex,\DeltaComplex')&=\overlapW (\varphi;\Delta,\Delta'),
\nonumber\\
&\approx
    e^{
        n
        \big(
            i\frac{\varphi}{2}
            -
            \frac{\pi(
                3 (\Delta+\Delta')^2
                +
                20\Delta\Delta'
                )
            }{256\mathcal{B} (\Delta+\Delta')}
            \varphi^2
            -
            \frac{\pi(\Delta-\Delta')^2}{16\mathcal{B} (\Delta+\Delta')}
        \big)
    },
\end{align}
where the  dependence on $\phi$ and $\phi'$ reduces to the phase-difference, $\varphi \equiv \phi'-\phi\in[-\pi,\pi)$.  This reduction makes  $\overlapW$ \emph{block-circulant} in the phase coordinates, and $\overlapW$ is a  Gaussian localised near $\varphi=0$.  

The Hamiltonian function is also block-circulant in $\varphi$, 
\begin{align} \label{eqn:ham_phase}
\hamH (\DeltaComplex,\DeltaComplex')&=\hamH (\varphi;\Delta,\Delta'),\nonumber\\
&= \big(\mathsf{L}(\Delta_+,\Delta_-)+O(\varphi)\big)\, \overlapW(\varphi;\Delta,\Delta'),
\end{align}
where  $\Delta_{\pm}=(\Delta\pm\Delta')/2$ and the leading, \mbox{$\varphi$-independent} term in $\hamH /\overlapW$ is
\begin{align}
\mathsf{L}(\Delta_+,\Delta_-)&=\tfrac{n}{2 \BW}\Big( \Delta_+^2\big(\!-\!\tfrac{1}{2}+\ln(2\BW/\Delta_+)\!-\!\lambda \ln^2(2\BW/\Delta_+)\big)\nonumber\\
&\quad\quad\quad+\Delta_-^2 \big(\tfrac{2}3-\ln(2\BW/\Delta_+)\big)\Big).
\label{eqn:L}
\end{align}

\Cref{fig:Hdiag} shows the `BCS potential', {$\mathsf{V}_{\rm BCS}(\Delta)\equiv\hamH (0;\Delta,\Delta)\approx \mathsf{L}(\Delta,0)$}, along the main diagonal $\Delta'=\Delta$, attaining a minimum \mbox{$-E_{\rm min}=-n\BW e^{-2/\lambda}$} at \mbox{$\Delta=\DeltaBCS $}.  In mean-field theory, the surface  generated by rotating $\mathsf{V}_{\rm BCS}$ through $-\pi\leq \phi<\pi$ is the  ``Mexican hat potential"  \cite{ref:mypaper_JJquantisation,ref:collectiveHiggsMode&couplingCDW}, in which the gap relaxes to $\DeltaBCS $.

\begin{figure}[t]
\includegraphics{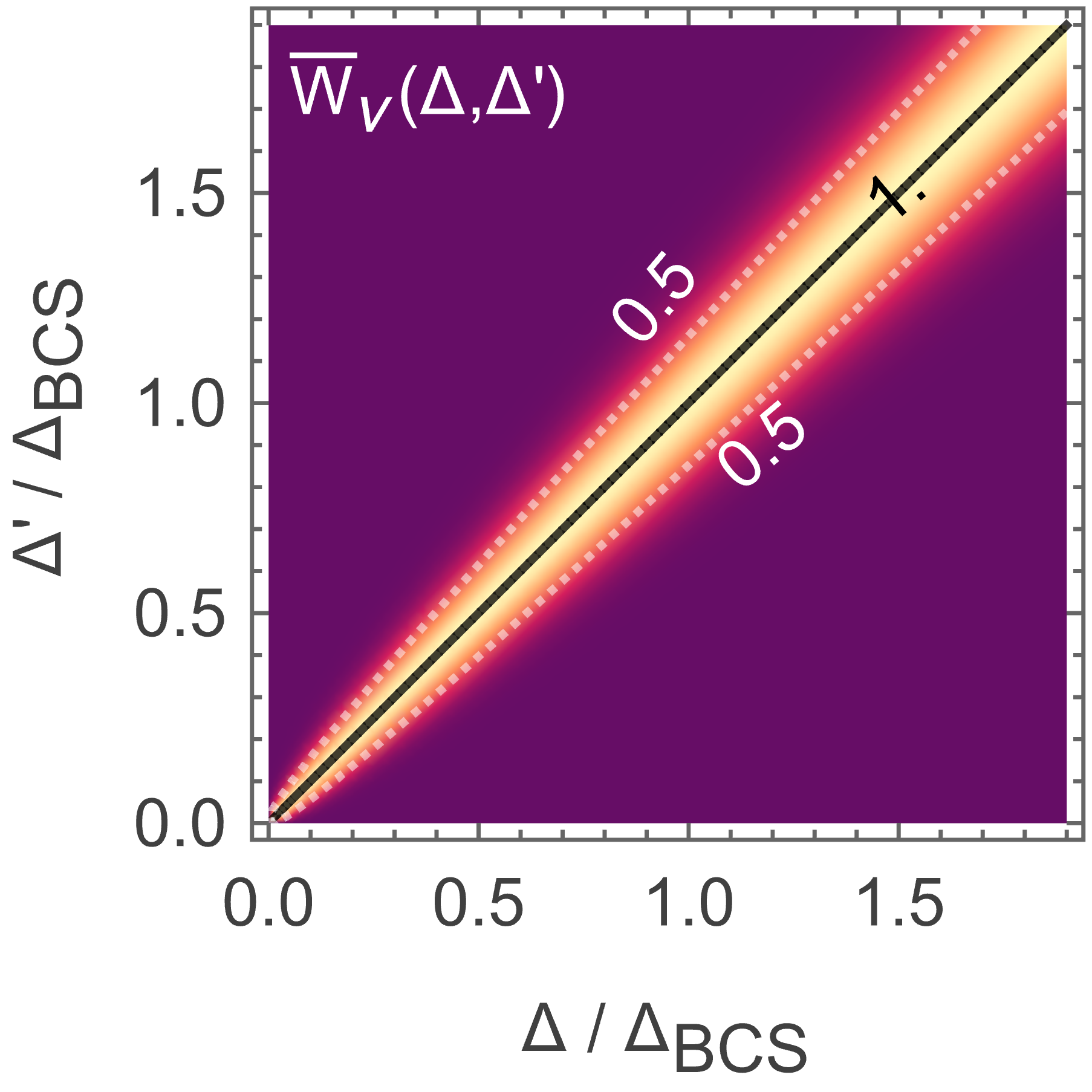}\hspace{-1mm}
\includegraphics{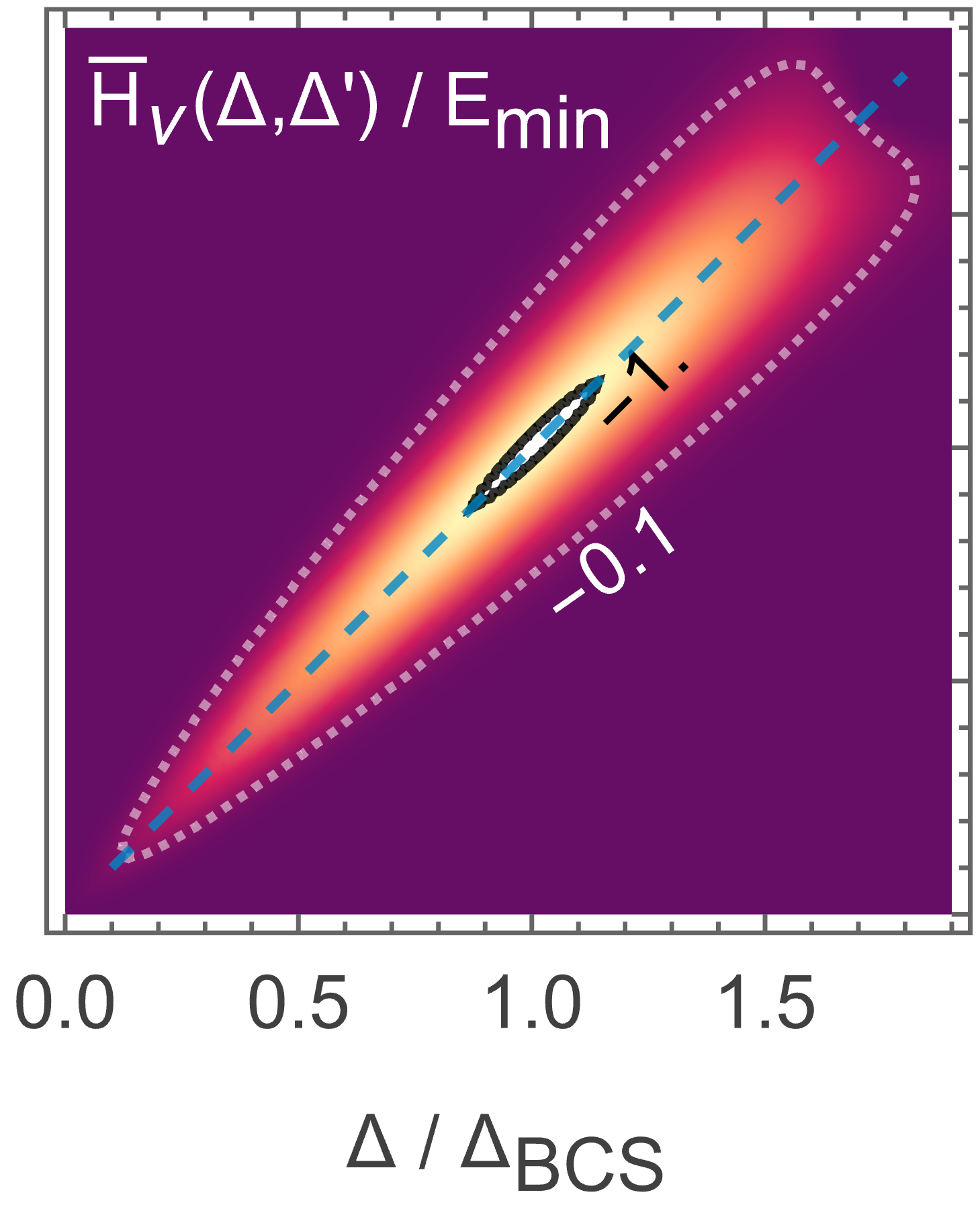}
\caption{
Normalised Fourier-basis overlap and Hamiltonian functions $\bar \overlapW_\nu$ and $\bar \hamH_\nu$ at half-filling, with  indicative contours  shown.    Along the diagonal $\Delta'=\Delta$ (dashed line),  $\bar \hamH_\nu(\Delta,\Delta)$ is qualitatively the same as shown in \cref{fig:Hdiag}, with a minimum  value of $-E_{\rm min}$  at $\Delta=\DeltaBCS$.  For illustrative purposes, we take  $\nu=0$, $n=3\times10^5$ and \mbox{$b=e^{1/\lambda}/2=10^3$}. 
}
\label{fig:WHplot}
\end{figure}

The block-circulant structure of $\overlapW$ and $\hamH$ allows simultaneous block-diagonalisation  using a discrete Fourier transform to replace the compact phase coordinate with a conserved island-charge quantum number $\nu\in\mathbb{Z}$, chosen so that $\nu=0$ corresponds to exact half-filling,
\begin{align}
 \overlapW_\nu(\Delta,\Delta')&={\medint{\int_{-\pi}^\pi}}d\varphi \,e^{-i (\nu+n/2) \varphi} \overlapW(\varphi;\Delta,\Delta'), \nonumber
 \\
 \hamH_\nu(\Delta,\Delta')&={\medint{\int_{-\pi}^\pi}}d\varphi \,e^{-i (\nu+n/2) \varphi} \hamH(\varphi;\Delta,\Delta')\nonumber.
\end{align}
Since $\overlapW$ and $ \hamH$ include a narrow Gaussian factor centred at $\varphi=0$, we extend the integration limits to $\pm\infty$ without significant error.  In addition,  $ \overlapW_\nu$ is unnormalised, which  follows because, for circulant overlap matrices, the (unitary) Fourier transform maps the \mbox{nonortho--normal} phase basis to an \mbox{ortho--nonnormal} charge basis \cite{ref:mypaper_JJquantisation}. For convenience we normalise in the charge basis by defining
 \begin{align}
\bar \overlapW_\nu(\Delta,\Delta')&= \overlapW_\nu(\Delta,\Delta')/\mathsf{N}_\nu(\Delta,\Delta'),\nonumber\\
&= 
\big(1+\Delta_-^2/(16\Delta_+^2)+ O(\nu^2)
\big)
e^{-\frac{n\pi\Delta_- ^2}{8\BW \Delta_+}}
   ,\label{eqn:Wnu} 
 \end{align}
 where $\mathsf{N}_\nu(\Delta,\Delta')
 =\sqrt{\overlapW_\nu(\Delta,\Delta)\overlapW_\nu(\Delta',\Delta')}$, so that $\bar \overlapW_\nu(\Delta,\Delta)=1$, and
  \begin{align}
\bar \hamH_\nu(\Delta,\Delta')&= \hamH_\nu(\Delta,\Delta')/\mathsf{N}_\nu(\Delta,\Delta')\label{eqn:Hnu},\nonumber\\
&=\big( \mathsf{L}(\Delta_+,\Delta_-)+O(\nu)\big) \,\bar\overlapW_\nu(\Delta,\Delta').
\end{align}
This normalisation leaves the spectrum invariant, and the integral eigenvalue  \cref{eqn:geneval}  reduces to
\begin{equation}
{\medint{\int
}} d\Delta'\, \bar \hamH_\nu(\Delta,\Delta')y_\nu(\Delta')\!=\!
E  \!{\medint{\int
}} d\Delta' \,\bar \overlapW_\nu(\Delta,\Delta')y_\nu(\Delta'),\label{eqn:evalprob}
\end{equation} 
where the conserved charge $\nu$ appears parametrically.

\Cref{fig:WHplot} shows $\bar\overlapW_\nu(\Delta,\Delta')$ and $\bar\hamH_\nu(\Delta,\Delta')$  at half-filling.   Qualitatively, these have support near the  diagonal  $\Delta=\Delta'$.  Along this main diagonal, $\bar\hamH_\nu(\Delta,\Delta)$ attains a minimum value  \mbox{$-E_{\rm min}$} at \mbox{$\Delta=\DeltaBCS $};  this behaviour for a fixed charge $\nu$ is essentially the same as that for a fixed phase, i.e.\ $\bar\hamH_\nu(\Delta,\Delta)\approx\mathsf{V}_{\rm BCS}(\Delta)$, shown in \cref{fig:Hdiag}.

\begin{figure}[t]
\includegraphics{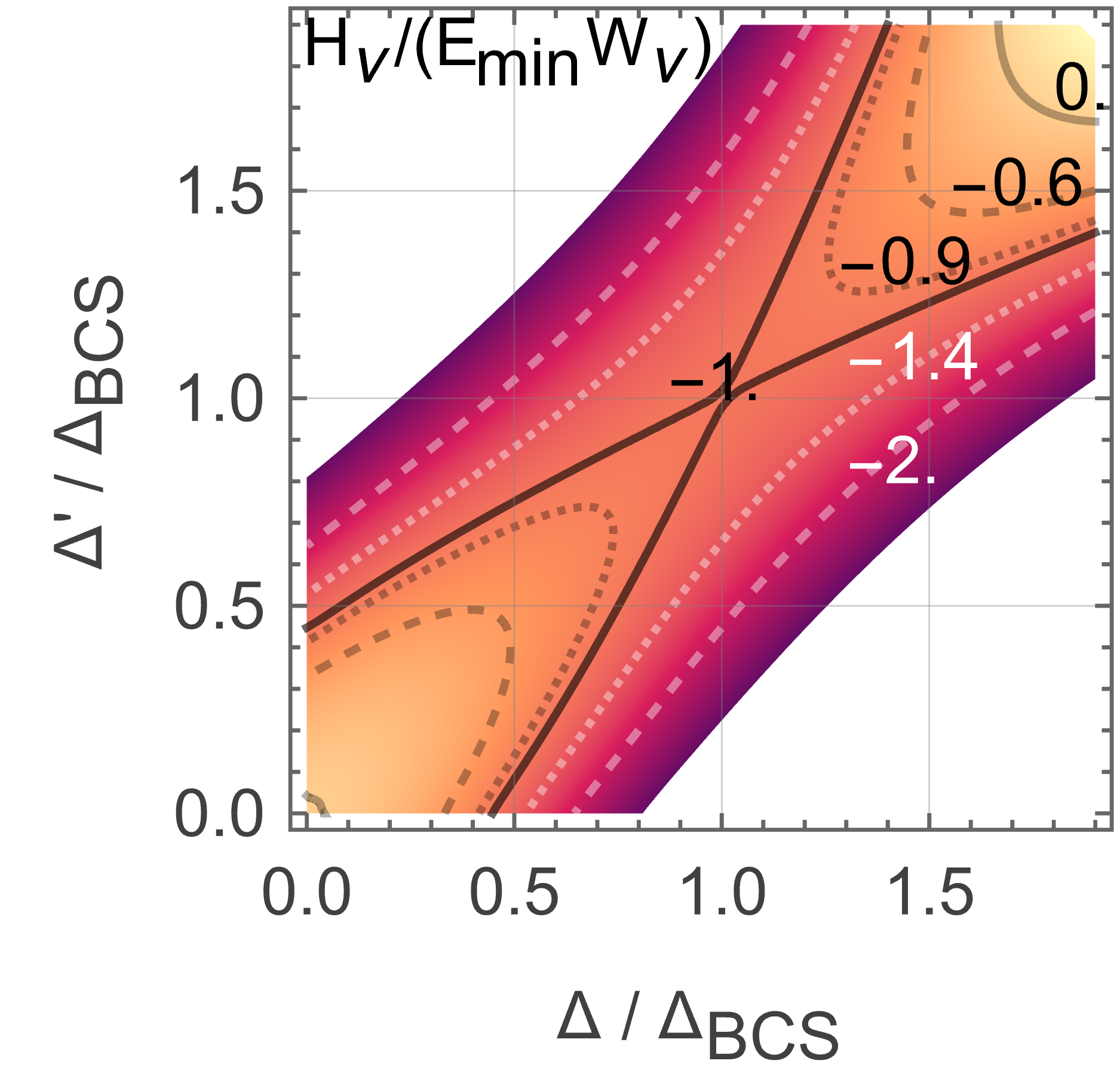}
\caption{
Illustrating the saddle-point structure in the ratio of the Hamiltonian and overlap functions $\hamH_\nu(\Delta,\Delta') / \overlapW_\nu(\Delta,\Delta')$, around $\Delta=\Delta'=\DeltaBCS$.  Parameters are as in \cref{fig:WHplot}, with  indicative contours  shown.  
}
\label{fig:HonWplot}
\end{figure}

Both $\bar\overlapW_\nu$ and $\bar\hamH_\nu$ exhibit a Gaussian decay as a function of the transverse coordinate \mbox{$\Delta_-$}. 
 These `off-diagonal' Hamiltonian terms capture the kinetics of the gap mode, which we formalise below. 
In addition, we note that the ratio \mbox{$\hamH_\nu / \overlapW_\nu=\mathsf{L}+O(\nu)$}  exhibits a  saddle-point  at \mbox{$\Delta=\Delta'=\DeltaBCS $,} as  illustrated in  \cref{fig:HonWplot}.  Analytically, we expand $\mathsf{L}$ in \cref{eqn:L}  to quadratic order   around \mbox{$\Delta_+=\DeltaBCS$} and $
\Delta_-=0$, to find
\begin{align}
\!\mathsf{L}/{E_{\rm min}}=&
\big(2-2/\ln(2b)\big)\,(\Delta_+/\DeltaBCS-1 )^2-1
\nonumber\\
&-\!\big(2\ln(2 b)-\tfrac{4}{3}\big)\,(\Delta_-/\DeltaBCS )^2+O(\Delta_\pm^3),\label{eqn:LonEmin}
\end{align} 
where $b=\BW/\DeltaBCS=e^{1/\lambda}/2$ is a dimensionless parameter depending on microscopic quantities.  For typical metals, \mbox{$\BW\gtrsim1$ eV} and \mbox{$\DeltaBCS\!<\! 1$ meV} \cite{ref:Al_gap}, so $b\gtrsim10^3$. 
In explicit electron-phonon interaction models the `ultraviolet' cutoff $\BW$ is often replaced by a characteristic phonon frequency $\bar{\omega}_{\rm ph}\!\sim\! 0.1\text{ eV}$ \cite{DYNES1972615}, so  {$b\!\rightarrow b_{\rm ph}\!=\!\bar{\omega}_{\rm ph}/\DeltaBCS\gtrsim 10^2$}.

We now extract effective spring and mass constants, $\kappa_\higgs$ and $m_\higgs$, from $\mathsf{L}$ in \cref{eqn:LonEmin}, by comparing it  with  the Hamiltonian function for a 1D particle of mass $m$ in a potential ${\rm v}(x)$, 
 with Hamiltonian $\hat h={\rm v}(\hat x)+\hat p^2/(2m)$.  
 We express $\hat h$  in a normalised, nonorthogonal  basis of  $\gamma$-localised Gaussian wavepackets, \mbox{$\{ \ket{\psi_{\gamma,\tilde x}\rangle}\}_{\tilde x\in\mathbb{R}}$} with  \mbox{$\langle x|\psi_{\gamma,\tilde x}\rangle=e^{-(x-\tilde x)^2/(2\gamma^2)}/\sqrt{\pi^{1/2}\gamma}$},  and overlap function 
\begin{equation}
\overlapw(\tilde x,\tilde x')=\bra{\psi_{\gamma,\tilde x}}\psi_{\gamma,\tilde x'}\rangle=e^{-\tilde x_-^2/\gamma^2}\label{eqn:wsp},
\end{equation} where $\tilde x_\pm=(\tilde x\pm\tilde x')/2$.  For a harmonic potential \mbox{${\rm v}( x)=\kappa( x-x_0)^2/2-{\rm v}_0$} the Hamiltonian function is
\begin{align}
\hamh(\tilde x,\tilde x')&=\bra{\psi_{\gamma,\tilde x}}\hat{h}\ket{\psi_{\gamma,\tilde x'}} \nonumber,\\
&=\overlapw(\tilde x,\tilde x')\big(  \kappa (\tilde x_+- x_0)^2/2-{\rm v}_0-\tilde x_-^2/(2\gamma^4m)\nonumber\\
&\quad \quad\quad\quad\quad\quad{}+\kappa\gamma^2/4 +1/(4\gamma^2m) \big).
\label{eqn:hsp}
\end{align}
As for $\mathsf{L}$, we see that $\hamh/\overlapw$ also has a  characteristic saddle-point at $\tilde x_+=x_0$ and $\tilde x_-=0$ (i.e.\ at $\tilde x=\tilde x'=x_0$), with principal axes aligned along $\tilde x_\pm$.  The $\kappa$-dependent (potential) contribution to $\hamh/\overlapw$ depends only on $\tilde x_+$, while the mass-dependent (kinetic) contribution  depends only on $\tilde x_-$.  Conversely, given functional forms  for $\overlapw$ and $\hamh/\overlapw$ we can extract  $\gamma$,  $m$, and $\kappa$.

\begin{figure}[t]
\includegraphics{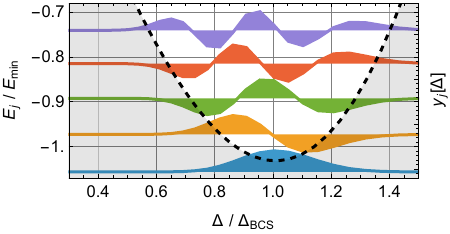}
\caption{
The lowest five Higgs eigenfunctions, ${\underline y}{}_{j}(\Delta)$, \mbox{$j=0,..,4$}, solved numerically on a discretised grid over $\Delta$, and superimposed on the BCS potential (dashed).  Eigenfunctions are displaced along the energy axis by their eigenenergy $E_j$.   Parameters are as in \cref{fig:WHplot}.   
}
\label{fig:evecs}
\end{figure}

To formally connect  to the superconducting gap dynamics, we treat the gap coordinate as a 1D particle in the BCS potential, with minimum at $\Delta=\DeltaBCS$.  With the correspondence $\Delta_\pm \leftrightarrow \tilde x_\pm$, we compare the  overlap functions $\overlapW_\nu$ and $\overlapw$ in \cref{eqn:Wnu,eqn:wsp}, and matching their Gaussian exponents yields 
\begin{equation}
\gamma_\higgs=\sqrt{{8\BW}\DeltaBCS /(n\pi)}=\sqrt{{8 b} /(n\pi)}\DeltaBCS.
\end{equation}
 Similarly, matching the coefficients of $\Delta_\pm^2$ in $\hamH_\nu/\overlapW_\nu\approx\mathsf{L}$ in \cref{eqn:LonEmin}, with those of $\tilde x_\pm^2$ in $ \hamh/\overlapw$ in \cref{eqn:hsp}, yields
\begin{align}
\kappa_\higgs/2&=\big(2-2/\ln(2b)\big){E_{\rm min}}/\DeltaBCS^2,\\
1/(2\gamma_\higgs^4m_\higgs)&=\big(2\ln(2 b)-\tfrac{4}{3}\big){E_{\rm min}}/\DeltaBCS^2.
\end{align}
Noting that  \mbox{$E_{\rm min}=n\DeltaBCS/(4b)$}, the Higgs frequency is
\begin{align}
\omega_\higgs=\sqrt{\kappa_\higgs/m_\higgs}
\approx \tfrac{8}{\pi} \sqrt{ \ln (2 b)\!-\!\tfrac{5}{3}}\DeltaBCS
\label{eqn:omegahiggs}.
\end{align}
For $b>10^2$,  we find $\omega_\higgs/\DeltaBCS>5$.

In addition, we expect zero-point gap fluctuations around $\DeltaBCS$ with standard deviation 
\begin{align}
\sigma_{\Delta}&=1/(\kappa_\higgs m_\higgs)^{1/4}
\approx   {\ln (2 b)}^{1/4}\gamma_\higgs.
\label{eqn:zpm}
\end{align}

The filled Fermi sea at $T=0$ is the quantum state at \mbox{$\Delta=0$} with $E=0$, shown in \cref{fig:Hdiag}, so bound states are those with $E<0$.  In the harmonic approximation, we  estimate the number of bound Higgs excitations to be
\begin{align}
N_{\rm bound}&\approx E_{\rm min}/\omega_\higgs
\approx  n \pi/\big(32\, b \sqrt{ \ln (2 b)-{5}/{3}} \big).\label{eqn:nmodes}
\end{align}
Qualitatively, for large system size $n$, when $N_{\rm bound}\gg1$, we anticipate that the lowest eigenmodes will be highly harmonic.  Conversely, when $N_{\rm bound}\sim1$, we expect significant anharmonicity in the low-energy spectrum.

We numerically validate the  harmonic analysis above by discretising the gap parameter $\Delta\rightarrow \Delta_i$, which corresponds to choosing a discretised set of gap-mode basis states $   \ket{\Psi_\nu(\Delta_i)}$. This transforms the integral eigenvalue  equation (\ref{eqn:evalprob}) into a  generalised matrix eigenvalue problem
$
\textsf{\textbf{H}}.{\underline y}=E\,  \textsf{\textbf{W}}.{\underline y},\label{eqn:matrix}
$
with  Hamiltonian and overlap matrices,  $\matH$ and $\matW$, which depend parametrically on $\nu$. 

For numerics,  we use a non-uniform grid over $\Delta$, chosen so that $|\matW_{i,i+1}|=|\bra{\Psi (\Delta_{i})}\Psi (\Delta_{i+1})\rangle |=1-\epsilon$, where $0<\epsilon\ll1$ controls the  overlap, independent of $i$.  For a finely discretised grid, where 
\mbox{$|\Delta_{i+1}-\Delta_i|\ll\gamma_\higgs$}, adjacent gap basis states are highly overlapped, and 
$\matW$  becomes increasingly ill-conditioned as  \mbox{$\epsilon\rightarrow0$}. 
 We find good numerical convergence for $\epsilon\gtrsim0.07$, but for  $\epsilon\lesssim0.05$  ill-conditioning   degrades numerical results.

\Cref{fig:evecs} shows eigenmodes $\underline y{}_j(\Delta)$ computed numerically, and displaced along the energy axis by their energy, $E_j$. Notably, the ground-state  energy is slightly below $-E_{\rm min}$, which is a consequence of the non-orthogonality of the gap-basis.  [This is reflected in the constants in the second line of \cref{eqn:hsp}, which raise the saddle-value of $\hamh$ above $-{\rm v}_0$.]   
The low-energy generalised eigenfunctions are qualitatively similar to harmonic oscillator modes, with the ground state  localised near \mbox{$\Delta=\DeltaBCS$}. For the parameters used in \cref{fig:evecs}, \cref{eqn:zpm} gives \mbox{$\sigma_\Delta/\DeltaBCS\approx0.15$},  consistent with the ground-state width shown.

When $N_{\rm bound}\gtrsim2$ we find that the first transition energy matches the Higgs frequency,  \mbox{$E_{10}\equiv E_1\!-\!E_0\approx \omega_\higgs$}, as expected.   For example, for the parameters used in \cref{fig:evecs} we find numerically \mbox{$E_{10}/E_{\rm min}=0.084$},  in  agreement with \mbox{$\omega_\higgs/E_{\rm min}=1/N_{\rm bound}=0.083$} from \cref{eqn:nmodes}.   

\begin{figure}[t]
\includegraphics{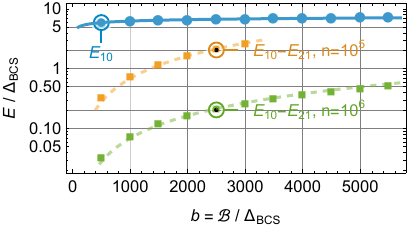}
\caption{
The first excitation transition energy \mbox{$E_{10}\equiv E_{1}-E_0$} computed numerically (circles) is in close agreement with the Higgs frequency $\omega_\higgs$ (solid line, \cref{eqn:omegahiggs}), as a function of the dimensionless bandwidth $b$; $E_{10}$ is essentially independent of system size $n$.  We also show the anharmonicity, \mbox{$\alpha_{210}=E_{10}-E_{21}$} (squares), for $n=10^5$ and $10^6$. Larger systems   are less anharmonic.   Dashed curves are the semi-analytic expression $15\,\alpha_{210}^{(3)}$, based on perturbative shifts.
}
\label{fig:spectrum}
\end{figure}

\Cref{fig:spectrum} shows $E_{10}$ computed numerically  (circles), and $\omega_\higgs$ from \cref{eqn:omegahiggs} (solid line), against the dimensionless  bandwidth, $b$.  The logarithmic dependence $\omega_\higgs/\DeltaBCS\approx\tfrac{8}{\pi}\sqrt{\ln(2b)}>5$ for small islands exceeds the long-wavelength prediction of $\omega_\higgs/\DeltaBCS\approx2$  \cite{ref:collectiveHiggsMode&couplingCDW,ref:Higgs_dynamics_Floquet}. 
We understand this difference  arises from the quasi-zero-dimensional limit we consider, where the system size is smaller than the superconducing coherence length.

\Cref{fig:spectrum} also shows the `quantum' anharmonicity, \mbox{$\alpha_{210}=E_{10}-E_{21}$} (squares) computed numerically for $n=10^5$ and $10^6$, showing frequency softening, and empirically scales  as \mbox{$\alpha_{210}\propto b\ln(2b)/n$}.  Larger systems have smaller anharmonicity, as expected, while $\alpha_{210}$ can be an appreciable fraction of $\omega_\higgs$ for small systems.  

The cubic potential contribution to  $\mathsf{L}$ in  \cref{eqn:L}
is \mbox{$\mathsf{V}^{(3)}_{\rm BCS}=E_{\rm min}\chi (\Delta_+/\DeltaBCS-1 )^3$}, where \mbox{$\chi=2/3-2/\ln(2b)$}, giving rise to second-order perturbative  energy shifts.  These  contribute an anharmonic shift  
$\alpha_{210}^{(3)}=\tfrac{15\,\sigma_\Delta^6}{2\,\omega_H}\big(\tfrac{E_{\rm min}\chi}{\DeltaBCS^3}\big)^2
\approx 40\,b\ln(2b)/(3n\pi^2)$.  The functional dependence on $n$ and $b$ in $\alpha_{210}^{(3)}$ agrees with the numerical behaviour of $\alpha_{210}$, but quantitatively underestimates it  by a factor $\approx15$.  We attribute this  to (a) higher-order potential contributions, and (b) the dependence of  $m_\higgs$  on $\Delta_+$ which localises excited eigenmodes,  increasing the wavefunction overlaps that contribute to  anharmonic perturbations.

Using aluminium as an example, {$\DeltaBCS= 340\,\mu{\rm eV}= 82\,(2\pi\,{\rm GHz})$} and \mbox{$\mathcal{B}=11\,{\rm eV}$} \cite{ref:Al_band}, so \mbox{$b\approx 3\!\times \!10^{4}$}.  \Cref{eqn:omegahiggs} predicts a Higgs  frequency of $\omega_\higgs\approx 7.8 \DeltaBCS$, so \mbox{$
\omega_\higgs
\approx 640\,({2\pi\,\rm GHz})$}.  For \mbox{$n=10^7$}, we compute  \mbox{$\alpha_{210}\approx0.35\DeltaBCS=28\,({2\pi\,\rm GHz})$}.  Aluminium has a 4-atom face-centred cubic unit cell of side length \mbox{$l_{c}=0.4$} nm, so at half-filling \mbox{$n=10^7$} would correspond  to a cubic island of side length \mbox{$l=n^{1/3}l_c/2\approx43$ nm}.  This suggests promising experimental directions for observing a superconducting Higgs transition in the near-THz band, that, for moderately small isolated islands, would have an appreciable quantum anharmonicity.

To conclude, the  analysis presented here extends the projective approach to circuit quantisation \cite{ref:Mizel2024,ref:mypaper_JJquantisation} to include anharmonic Higgs excitations, expanding  superconducting circuit theory beyond the ground-state Higgs-mode manifold.  We have focused  on the physics of a small island, whose Hilbert space is parameterised by a single complex coordinate $\DeltaComplex$, but this can be expanded to spatially delocalised systems.   Our results suggest an exciting  direction in superconducting electronics that  uses the intrinsic anharmonicity of superconducting Higgs modes of small  islands to encode and manipulate quantum information, akin to the two-band superconductor proposed by \citet{ref:Leggett1966}.  With near-THz transition frequencies, and   anharmonicities \mbox{$\gtrsim10$ GHz} in nanometer-scale electronics,  compact superconducting qubit systems may be feasible, with correspondingly high operating frequencies and  temperatures.  Key to developing this direction is understanding the relaxation mechanisms and lifetimes of Higgs modes, and the complementary task of practical  control of such devices. 

\begin{acknowledgments}
The authors  acknowledge funding from the Australian Research Council Centre of Excellence for Engineered Quantum Systems CE170100009, and thank S.\ Barrett and A.\ Doherty for formative discussions.
\end{acknowledgments}

\bibliographystyle{apsrev4-2}
\bibliography{reference.bib}

\end{document}